\documentclass[baaa]{baaa}

\usepackage[pdftex]{hyperref}
\usepackage{subfigure}
\usepackage{natbib}
\usepackage{helvet,soul}
\usepackage[font=small]{caption}

\contriblanguage{1}

\contribtype{1}

\thematicarea{2}

\received{\ldots}
\accepted{\ldots}

\title{Dynamical fractalization of ring systems}

\titlerunning{Dynamical fractalization of ring systems}

\author{
J. Ruta\inst{1,2}, N.E. Grandi\inst{2,3} \& T. Canavesi
}

\authorrunning{Ruta et al.}

\contact{rutajere@gmail.com}

\institute{
Facultad de Ciencias Astron\'omicas y Geof{\'\i}sicas, UNLP, Argentina
\and
Instituto de F{\'\i}sica La Plata, CONICET--UNLP, Argentina
\and
Departamento de F{\'\i}sica Dr. Emil Bose, Facultad de Ciencias Exactas, UNLP, Argentina
}

\resumen{Desarrollamos un modelo efectivo para describir la dinámica de un sistema de partículas que se mueven en órbitas circulares en torno a una masa central, considerando el límite continuo de la distribución angular, para obtener las configuraciones estables para distintos parámetros iniciales. Calculamos la dimensión fractal resultante y la comparamos con la dimensión de un conjunto de anillos de Cantor aleatorios.}

\abstract{We develop an effective model to describe the dynamics of a system of particle moving in circular configurations around a central mass, by considering the continuum limit of the angular distribution, to obtain the stable configurations for different initial parameters. We compute the resulting fractal dimension and compare it with that of a statistical Cantor ring system.}


\begin{document}

\maketitle
\section{Intoduction}
\label{intro}
Given a set of points in three-dimensional space, we can define its \emph{box counting dimension} as follows. First, we divide the space in a cubic lattice made of cells of side $\ell$, and count the number ${\cal N}$  of such cells which have a non-vanishing intersection with the set. Then, we scale the lattice parameter $\ell$ and fit the dimension $d$ according to the formula ${\cal N}\sim \ell^{-d}$. To the scope of the present note, it will be enough to define a \emph{fractal} as a set of points whose box counting dimension $d$ is not an integer. Fractals appear everywhere in nature, from the ramifications of the circulatory and nervous systems in the human body, to the growth of plants and trees and the folding of the meanders of rivers.

The idea that planetary rings may have a fractal structure can be traced back to \cite{mandelbrot1982fractal} who, early in the history of fractal geometry, wondered whether the rings of Saturn correspond to some kind of statistical Cantor set. More recent works (see \cite{li2015edges} and \cite{malyarenko2017fractal}) measured the box coounting dimension of Saturn's rings on the pictures taken by the planetary missions Voyager 1 and 2 and  Cassini.

It is the aim of this work is to verify whether a gravitationally interacting $N$-particle system arranged in circular orbits around a central mass, representing  a ring system, organizes into a fractal structure. To this end, we develop the effective gravitational potential that governs the dynamics of the ring radii, and search for the resulting equilibrium configurations by minimizing it numerically. We then compute the fractal dimension of the resulting structure by making use of the box-counting method. Finally, we compare the obtained dimension with that a Cantor ring system.

\section{Mathematical model}

For this model, we start with an $N$-particle system interacting gravitationally, where each particle is described by a position vector \(\mathbf{x}_i \in \mathbb{R}^2\), with \(i = 1, \dots, N\). The potential energy of the system is then given by the standard Newtonian expression
\begin{equation}
    V=- \frac G2\sum_{i}^N    \sum_{j}^N\frac{ m_im_j}{|\mathbf{x}_i-\mathbf{x}_j|}.
\end{equation}
This gravitational potential can be expressed in polar coordinates $\mathbf{x}_i=(r_i\cos\theta_i,r_i\sin\theta_i)$ as
%
%
%
\begin{align}
V=-\frac G2\sum_{r_i\theta_i}^N\sum_{r_j\theta_j}^N\frac{\,m_im_j}{\sqrt{r_i^2+r_j^2-2r_ir_j\cos(\theta_i-\theta_j)}}.
\end{align}
where, as the $i$-th particle sits at $(r_i, \theta_i)$, we have used the values of these variables as the indices of the sums. Defining the angular distance between adjacent particles around radius $r_i$ as $\Delta\theta_i$, we can rewrite the above expression in the form
\begin{align}
V=-\frac G2\sum_{r_i\theta_i}^N\sum_{r_j\theta_j}^N\frac{m_i\tilde m_j}{\sqrt{r_i^2+r_j^2-2r_ir_j\cos(\theta_i-\theta_j)}}{\Delta\theta_j},
\end{align}
where $\tilde m_j=m_j/\Delta\theta_j$ is the angular mass density. Taking the continuuum limit $\Delta\theta_j\to 0$ and assuming that $\tilde m_j$ is independent of the angle $\theta_j$ and depends only on the radius $r_j$, we have
\begin{align}
V=-\frac{G}{2}\sum_{r_i\theta_i}^N\sum_{r_j}^{N_r}\int\frac{ m_i\tilde m_j}{\sqrt{r_i^2+r_j^2-2r_ir_j\cos(\theta_i-\theta_j)}}{\mathrm{d}\theta_j},
\end{align}
where $N_r$ is the number of different radii $r_i$. 
%
%
Performing the integral over the entire orbit, we get
\begin{align}
V=-2 G\sum_{r_i\theta_i}^N\sum_{r_j}^{N_r}
\frac{m_i\tilde m_j}{|r_i-r_j|}
\,K\!\left(\frac{-4r_ir_j}{(r_i\!-\!r_j)^2}\right),
\end{align}
where $K(k) = \int_0^{\pi/2} {\mathrm{d}\psi}\,/{\sqrt{1 - k^2 \mathrm{sen}^2 \psi}}$ is the complete elliptic integral of the first kind. We can now write $m_i=\tilde m_i\Delta\theta_i$ and then rearrange, to obtain
%
%
%
%
%
\begin{align}
V=-4\pi G\sum_{r_i}^{N_r}\sum_{r_j}^{N_r}
\frac{ \, \tilde m_i\tilde m_j}{|r_i-r_j|}
\,K\!\left(\frac{-4r_ir_j}{(r_i\!-\!r_j)^2}\right).
\end{align}
where we have replaced the sum over angular intervals at radius $r_i$ as $\sum_{\theta_i}^{N_\theta}\Delta \theta_i=2\pi$. We now need to make a hypothesis about the angular mass density $\tilde m_i$. If we assume that it is a function only of the radius, then we can write $\tilde m_i=q'^2(r_i)/2\pi$. This expression is completely general, the reason for the derivative form will become evident later; for now, let us just notice that the square ensures positive masses. This leads to
\begin{align}
V=-2G \sum_{r_i}^{N_r}\sum_{r_j}^{N_r}\frac{   q'^2(r_i)q'^2(r_j)}{|r_i-r_j|}\,K\!\left(\frac{-4r_ir_j}{(r_i\!-\!r_j)^2}\right),
\end{align}
This is the form of our effective interaction potential, governing the dynamics of the radii $r_i$. 

With the potential found above, we can write a Lagrangian in polar coordinates as
\begin{align}    
L&=\sum_{r_i\theta_i}^N m_i\bigg(\frac12\left(\dot r_i^2+r_i^2\dot\theta_i^2\right)+\frac{GM}{r_i}\bigg)+V
\end{align}
where a central mass $M$ has been added. In the kinetic term, we write $m_i=\tilde m_i\Delta \theta_i$, and we assume that the angular velocities of the particles $\dot\theta_i$ do not depend on the angle. With this, we obtain
%
%
%
\begin{align}
L&= \sum_{r_i}^{N_r}
  q'^2(r_i)\bigg(\frac12\left(\dot r_i^2+r_i^2\dot\theta_i^2\right)+\frac{GM}{r_i}\bigg)+V
\end{align}
where as before, the angular sum resulted in a factor of $2\pi$, and the angular mass density was replaced by $\tilde m_i=q'^2(r_i)/2\pi$.
This Lagrangian is invariant under independent rotations of each of the angles $\theta_i\to\theta_i+\delta_i$, so the corresponding angular momenta $l_i=q'^2(r_i)r^2_i\dot\theta_i$ are conserved. To incorporate this simplification into the dynamics, we  work out the Routhian, 
as
\begin{align}
R&= \sum_{r_i}^{N_r}  q'^2(r_i)\bigg(\frac12\dot r_i^2-\frac{l_i^2}{2q'^4(r_i)r_i^2}+\frac{GM}{r_i}\bigg)+V
\end{align}
%
The kinetic term is not canonical, but if we redefine our canonical coordinates as $q_i=q(r_i)$, we get
\begin{align}
R&= \sum_{r_i}^{N_r} \bigg(\frac12\dot q_i^2-\frac{l_i^2}{2q'^2(r_i)r(q_i)^2}+\frac{q'^2(r_i)GM}{r(q_i)}\bigg)+V
\end{align}
where $r(q)$ is the inverse function of $q(r)$ (incidentally, here we see the reason why we defined $m_i=q'^2(r_i)$). From this, we can obtain the effective dynamics for the orbital radii.

Different models correspond to different choices of the function $q(r)$. For example, if we assume that the function $\tilde m_i$ is independent of $r$, this implies that $q'(r_i)=\rho$ is constant, so that $q(r_i)=\rho r_i$ and therefore $r(q_i)=q_i/\rho$. With this, we obtain the simplest possible model
\begin{align}
R&=\sum_{r_i}^{N_r}  \left(\frac12\dot q_i^2-\frac{l_i^2}{2q_i^2}+\frac{\rho^3 GM}{q_i}
\right)
+\nonumber\\&~~~~~~~~~~~~~+\sum_{r_i}^{N_r}
\sum_{r_j}^{N_r}\frac{2 G\,\rho^5}{|q_i-q_j|}\,K\left(\frac{-4q_iq_j}{(q_i\!-\!q_j)^2} 
\right).
\label{eq:ruthiano modelo simple}
\end{align}
Notice that in this model, the total mass of a circumference at any radius takes the same value $2\pi\rho^2$, so the outer rings are more dilute than the inner ones. 
%
%
%
This gives us the following effective gravitational potential for the radii
\begin{align}
    V_{\sf eff}= &
    \sum_{r_i}^{N_r} \!\left(\!\frac{l_i^2}{2q_i^2}-\frac{\rho^3GM}{q_i}
    -
    \sum_{r_j}^{N_r}\!\frac{2 G\,\rho^5}{|q_i-q_j|}K\!\left(\frac{-4q_iq_j}{(q_i\!-\!q_j)^2}\right)
    \right)
\end{align} 
To obtain the radii of the ring system, we need to minimize this potential for a given set of angular momenta.

\section{Numerical results}
We developed  {\tt Python} and {\tt Mathematica} codes to minimize the effective potential for a system of $N = 150$ particles, with $\rho = 10$ and a central mass of $GM = 3.5 \times 10^{4}$. 

To initialize the system, we sample a set of angular momenta from a normal distribution with different mean values $\bar l$ and standard deviations $\Delta l$. In a subset of our numerical experiments, we motivated the parameters from a physical perspective, as follows: we consider a satellite that exceeds the Roche limit $R=r_{\sf{sat}}({2M}/{m_{\sf{sat}}})^{1/3}$ 
%
%
where $r_{\sf{sat}}$ is the size of the satellite and $m_{\sf{sat}}$ is its mass. The mean value of the angular momentum $\bar l$ is then calculated as that of a mass $m_{\sf sat}$ in a circular orbit around $M$ at radious $R$. The standard deviation $\Delta l$ on the other hand is approximated as the angular momentum of a mass $m_{\sf sat}$ rotating at radious $r_{\sf sat}$ around the satellite centre with a tidally locked revolution period. 

\begin{figure}[!t]
\centering
\includegraphics[width=\columnwidth]{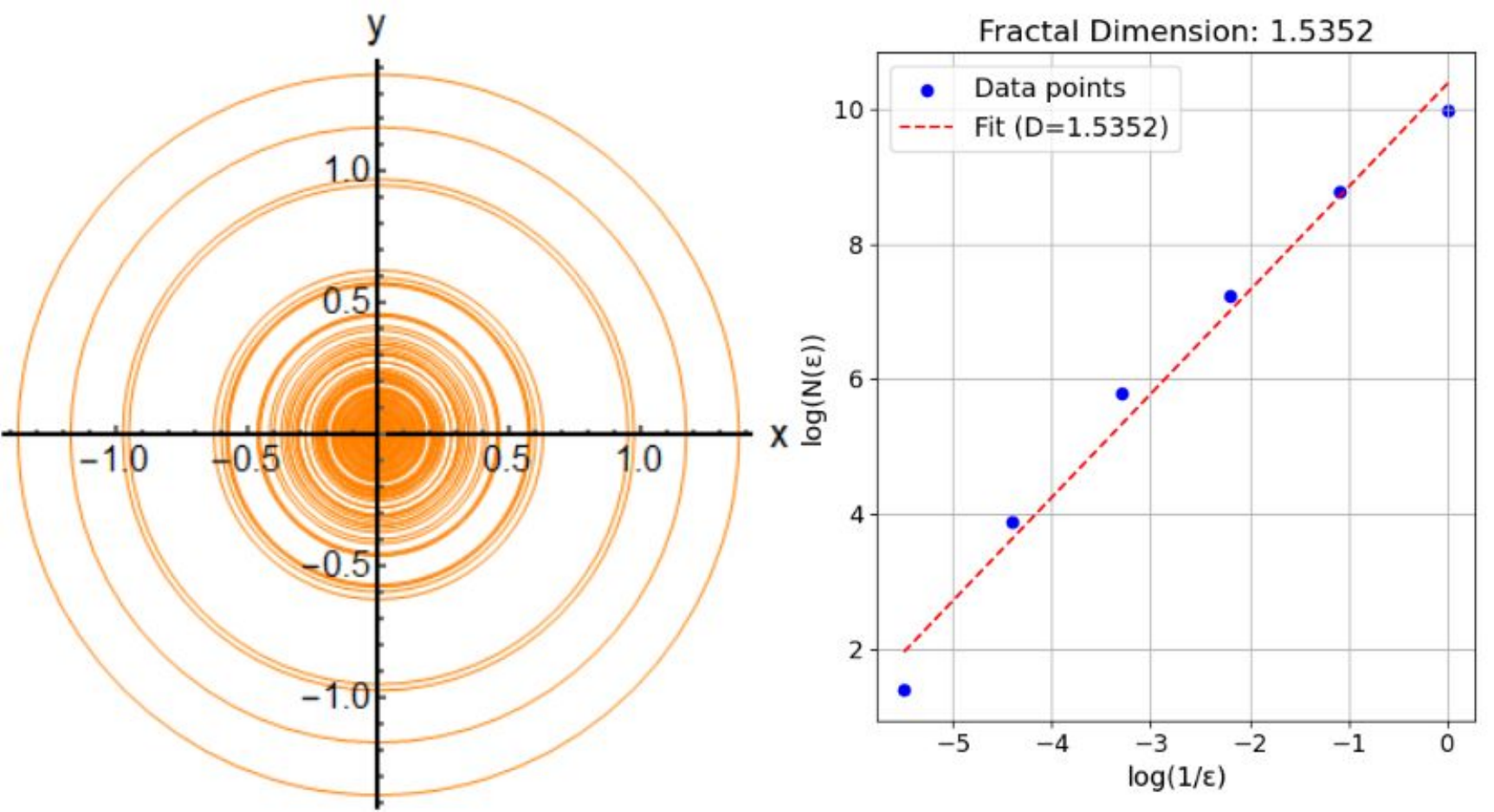}
\caption{Initial angular momenta drawn from a normal distribution with $\bar l = 1000$, $\Delta l=50$.}
\label{1000_50}
\end{figure}

\begin{figure}[!t]
\centering
\includegraphics[width=\columnwidth]{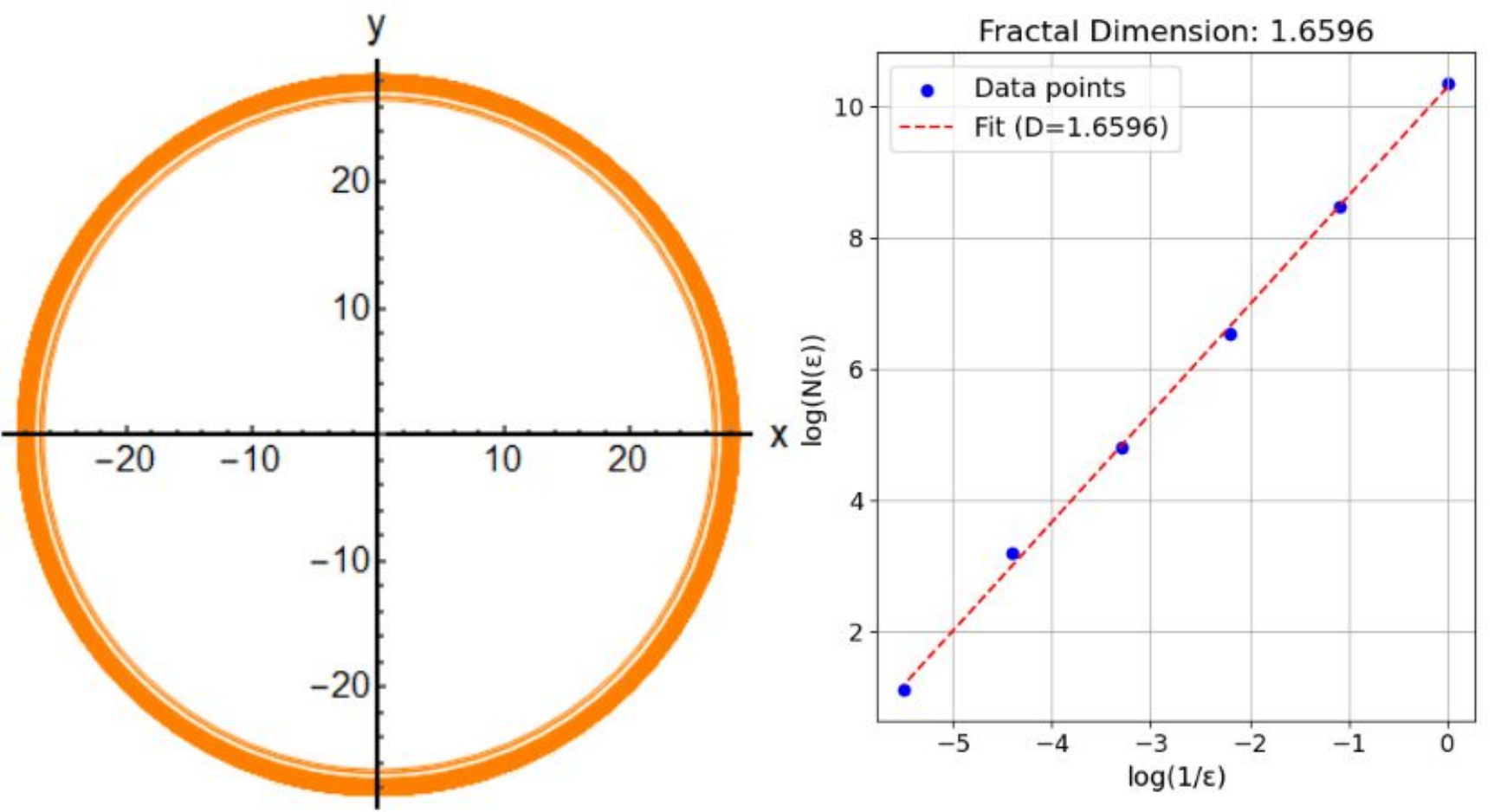}
\caption{Calculated using the Roche limit with $r_{\mathrm{sat}} = 1$ and $m_{\mathrm{sat}} = 1500$.}
\label{1000_200111}
\end{figure}

\begin{figure}[!t]
\centering
\includegraphics[width=\columnwidth]{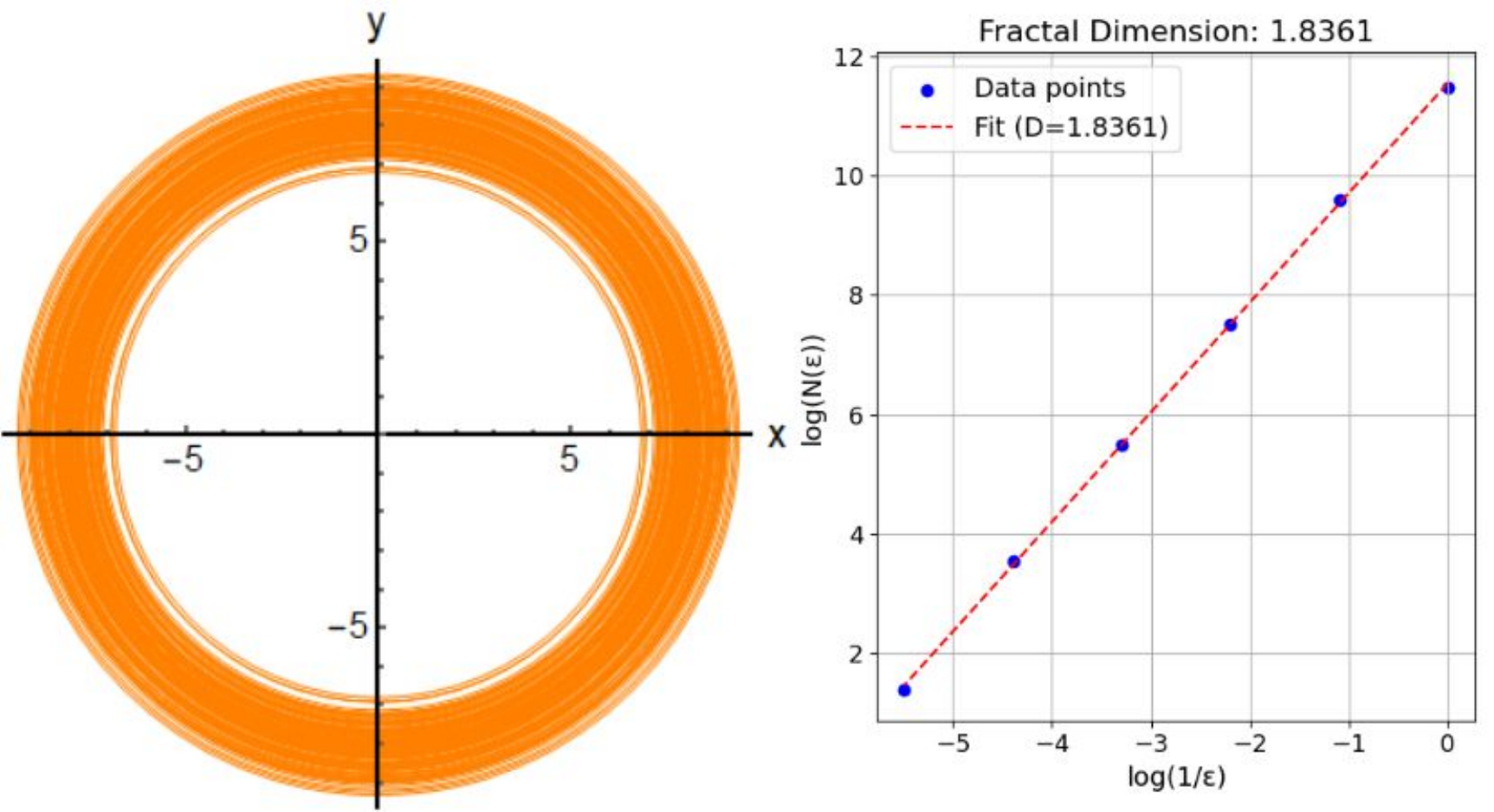} 
\caption{Calculated using the Roche limit with $r_{\mathrm{sat}} = 0.01$ and $m_{\mathrm{sat}} = 1500$.}
\label{1000_20011}
\end{figure}

\begin{figure}[!t]
\centering 
\includegraphics[width=\columnwidth]{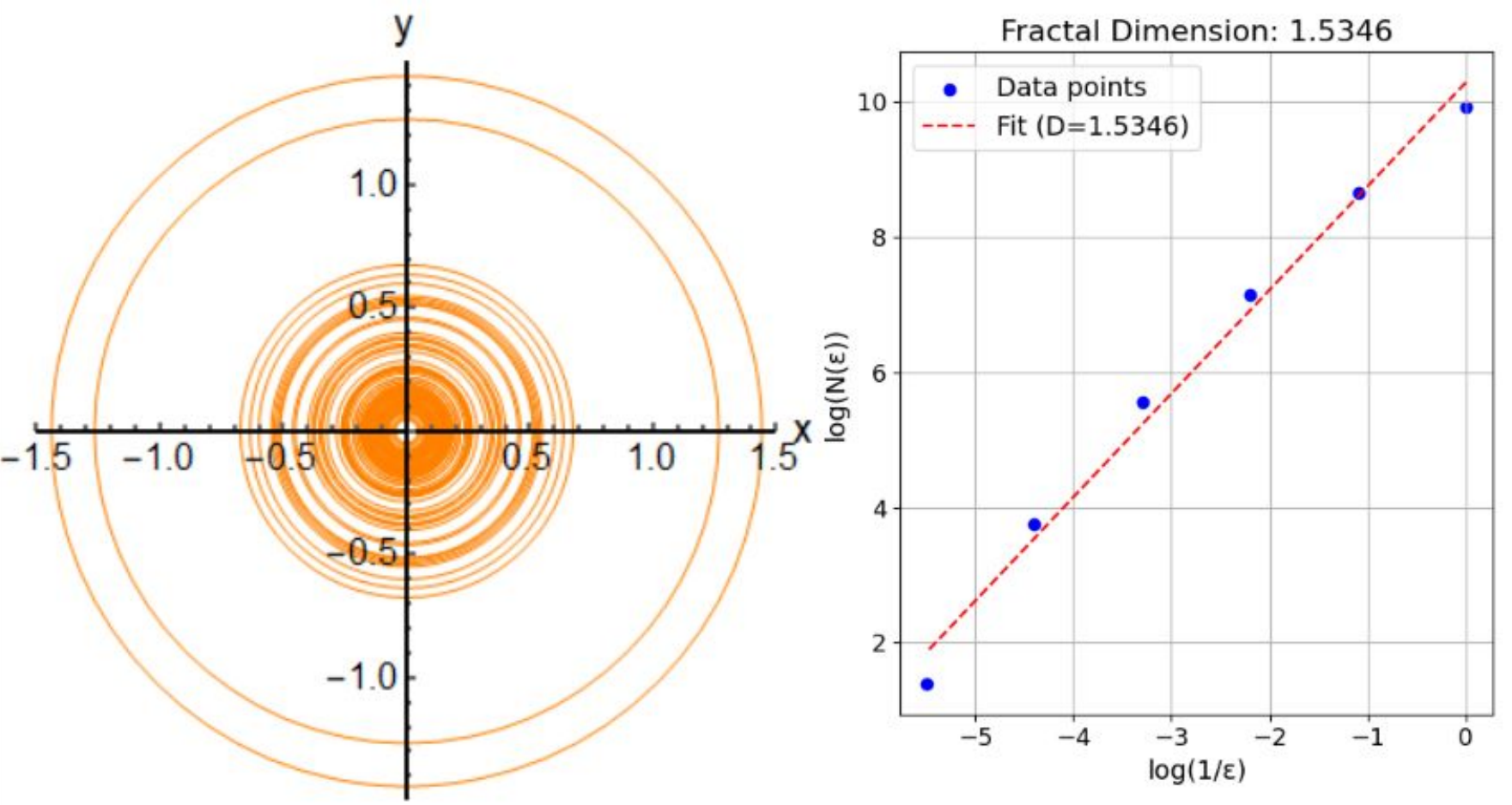}
\caption{Calculated using the Roche limit with $r_{\mathrm{sat}} = 0.0001$ and $m_{\mathrm{sat}} = 1500$.}
\label{1000_2001}
\end{figure}

The results from the numerical implementation are presented in Figs.~\ref{1000_50} and \ref{1000_2001}. 
Figs.~\ref{1000_200111} to \ref{1000_2001} correspond to a subset of simulations obtained 
from the calculation of the Roche limit. 
Representative examples of the simulations are shown for different parameter choices.  
Each figure consists of two panels: 
the left panel displays the final ring configuration,  
whereas the right one shows its box-counting dimension. 

\begin{figure}[!t]
\centering
\includegraphics[width=\columnwidth]{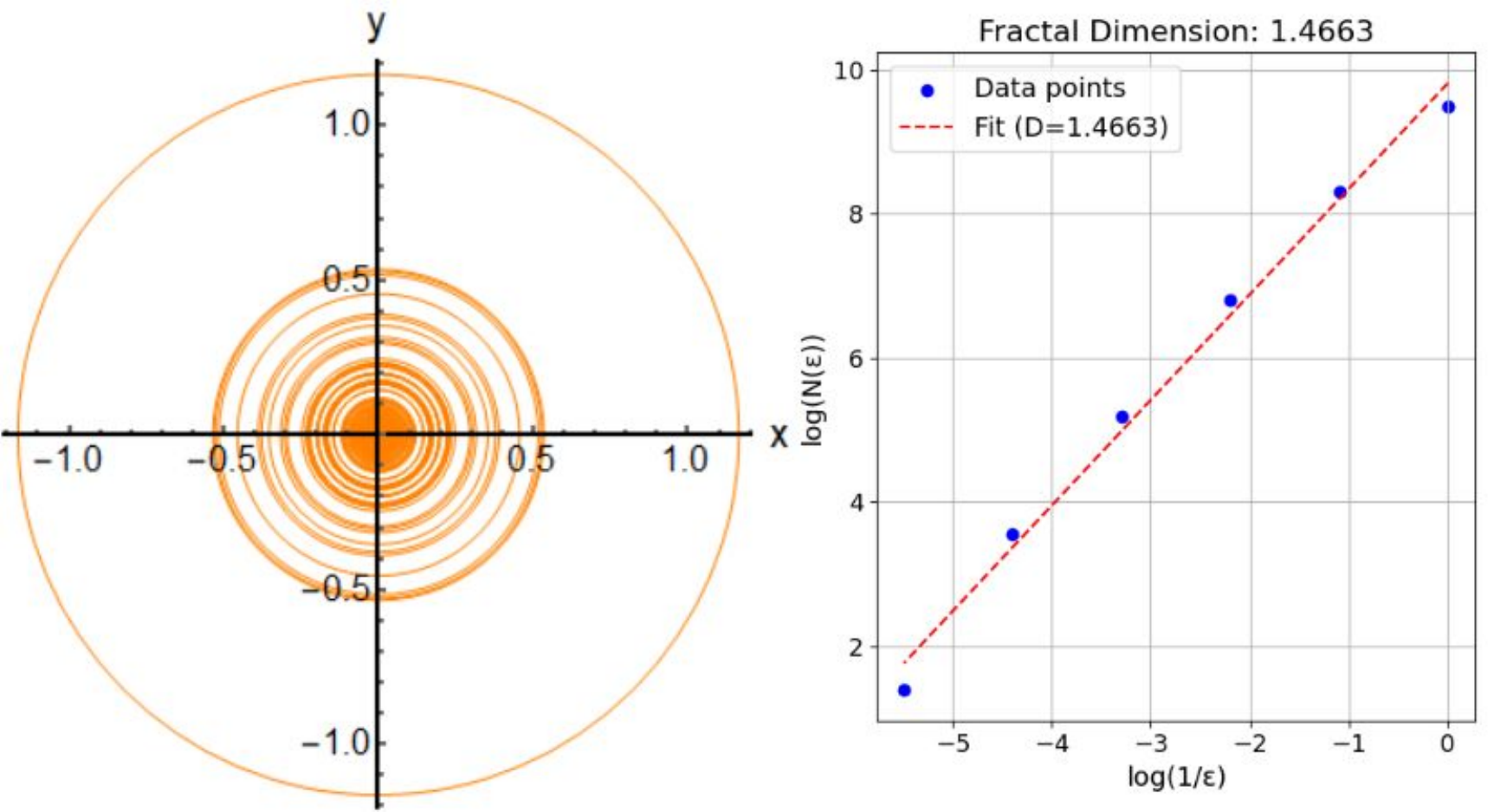}
\caption{Initial angular momenta drawn from a normal distribution with $\bar l = 1000$, $\Delta l=100$.}
\label{1000_100}
\end{figure}

\begin{figure}[!t]
\centering
\includegraphics[width=\columnwidth]{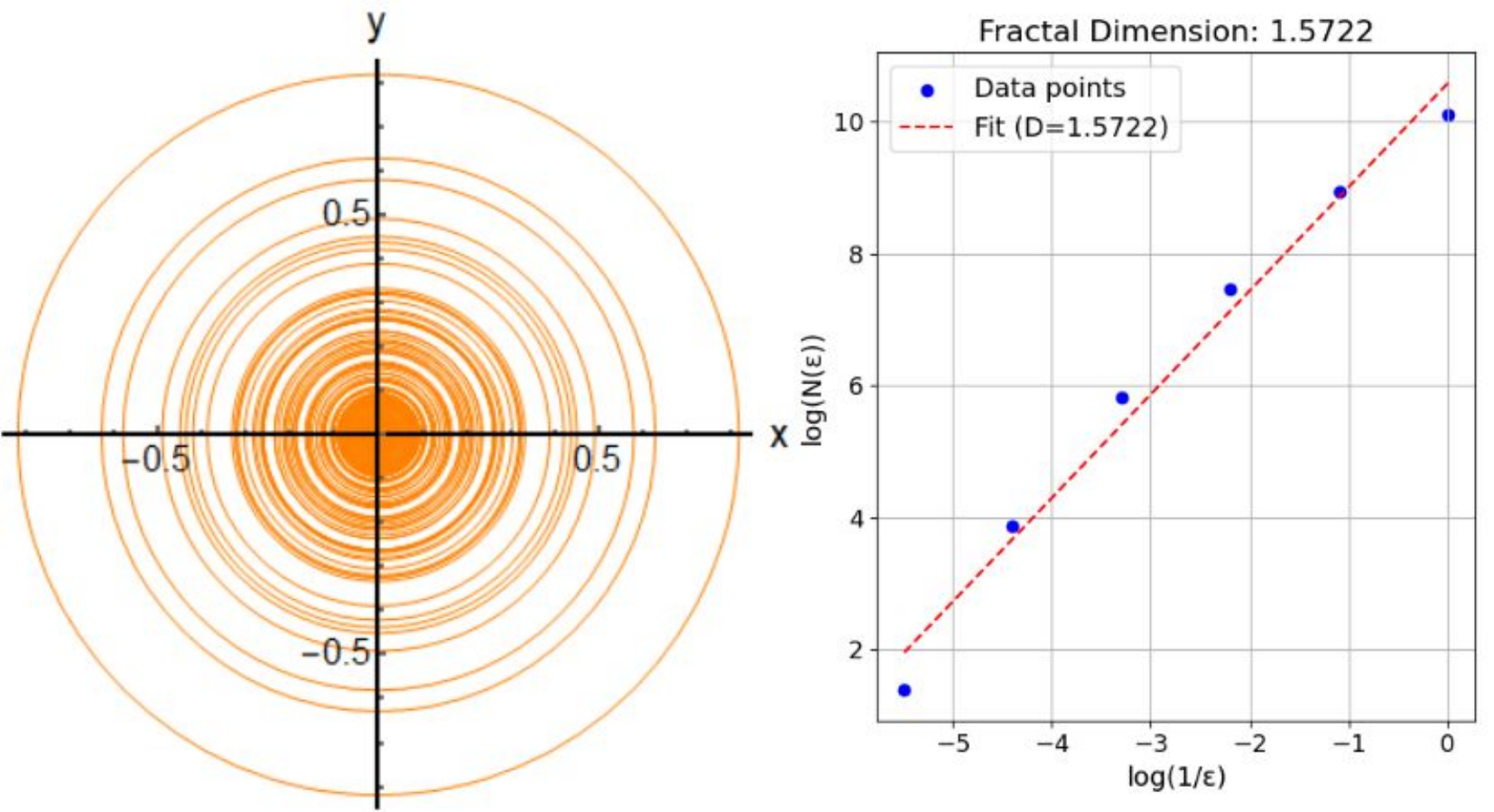}
\caption{Initial angular momenta drawn from a normal distribution with $\bar l = 1000$, $\Delta l=200$.}
\label{1000_200}
\end{figure}

The resulting dimensions lie between $1.45$ and $1.85$, with those corresponding to a satellite in its Roche limit in the upper end, while the ones generated from a gaussian distribution of angular momenta in the lower one.

\section{Discussion and outlook}

The obtained fractal dimensions can be compared with that a system of \emph{random Cantor rings}. It is generated as follows: (i) A one dimensional segment is cut in three pieces whose lentgh is sampled from a uniform distribution. (ii) The middle piece is removed and the random process is repeated in the remaining two components. (iii) After a given number of iterations (eight in our case), the set is revolved around a point exterior to the original segment, to generate a ring system. 
The resulting square counting dimension can be then calculated, ranging between $1.6$ and $1.7$. It is noteworthy that, for a wide range of parameters, this is very close to the fractal dimension of our dynamically constructed ring system.

An important developement of the present approach would be to investigate the role of the function $q(r)$, that we have chosen by hand in our previous calculations. As a first step, we plan to  explore a more general class of models, with the objetive of quantifying  the dependence of the resulting box counting dimension for the ring system on the chosen function $q(r)$. 
It is clear to us that this function has no direct physical meaning, and thus a more complete approach would entail to minimize the energy of the stationary configuration with respect to $q(r)$ in a self-consistent way. A possible way to proceed would be to expand $q(r)$ in powers of $r$ and treat the higher coefficients as perturbations, to evaluate its impact on the stability of the system. 

\begin{acknowledgement}
We are grateful to Daniel Carpintero and Milva Orsaria for discussions and helpful comments regarding the effective potential. This work was part of a MSc thesis to obtain an Astronomy degree at La Plata University. N.E.G. was partially supported by CONICET grant PIP2023-11220220100262CO and UNLP grant 2022-11/X931. J. R. was partially supported by an EVC-CIN fellowship. 
\end{acknowledgement}
\newpage


\bibliographystyle{baaa}
\small
\bibliography{bibliografia}
 
\end{document}